Determination of the energy of the Dzyaloshinskii-Moriya interaction in [Co/Pd(111)]$_5$ superlattices with different Co thickness by micromagnetic simulations of labyrinth domain structures.

A.G. Kozlov[1], A.G. Kolesnikov[1], M.E. Stebliy[1], and A.V. Davydenko[1,*]

[1]Laboratory of Thin Film Technologies, School of Natural Sciences, Far Eastern Federal University, Vladivostok, 690950, Russian Federation

Abstract

Determination of the energy of Dzyaloshinskii-Moriya interaction along with a definition of the basic magnetic characteristics in ferromagnetic/nonmagnetic multilayered systems are both required for the construction of a magnetic skyrmion recording medium. A method for estimating the energy of the effective Dzyaloshinskii-Moriya interaction which compared the periodicities of micromagnetically simulated and experimentally measured demagnetized domain structures was shown in a current paper. Symmetric epitaxial [Co/Pd(111)]$_5$ superlattices with Co layers of varying thickness were used as the system for investigation. The structural and magnetic properties of epitaxial [Co($d_{Co}$)/Pd]$_5$ superlattices with different Co layers thicknesses were comprehensively investigated. The dependence of the energy of effective Dzyaloshinskii-Moriya interaction on the thickness of the Co layers in the [Co($d_{Co}$)/Pd]$_5$ multilayered structures was determined. The relationship between Dzyaloshinskii-Moriya interaction and asymmetry of the strains between the bottom Pd/Co and top Co/Pd interfaces was discussed. The simulation parameters and the demagnetization approach prior to measuring the magnetic structure influenced the obtained results. Necessity of setting all the layers in micromagnetic simulations was established. The significant influence of interlayer dipolar coupling on the periodicity of simulated labyrinth domain structures was also confirmed.

I. INTRODUCTION

The Dzyaloshinskii-Moriya interaction (DMI) became one of the most intriguing phenomena in modern nanomagnetism. The necessary conditions for the existence of DMI in materials are strong spin-orbit coupling and a lack of inversion symmetry [1, 2]. These requirements may be fulfilled in the interfaces between ferromagnetic (FM) and nonmagnetic (NM) layers. Hence, interface DMI may be stabilized in ultrathin NM/FM/NM structures [3, 4]. It was reported that the DMI is rather strong at the interfaces between FM and heavy metals (HM) due to strong spin-orbit coupling in HM [5]. Additive DMI may be obtained in the HM1/FM/HM2 systems with an appropriate choice of HM layers [6, 7]. Strong interface DMI with perpendicular magnetic anisotropy (PMA) stabilize homochiral Neel domain walls (DWs) [8] and Neel skyrmions [9] which may be efficiently displaced by current pulses due to the spin-orbit torque (SOT) effect. Different concepts of racetrack memories based on DWs and skyrmions have already been introduced [10, 11]. An important step towards working racetrack memory was switching researchers' attention to multilayered structures or superlattices [HM1/FM/HM2]$_N$ [6, 7]. On the one hand, PMA and DMI are conserved in such magnetic superlattices due to a large number of interfaces. On the other hand, the overall thickness of the structures may be sufficiently increased which makes the skyrmions thermally stable. Moreover, growing dipolar interaction demagnetizes the structures and facilitates the formation of skyrmions.

Construction of the chiral magnetic systems with the required characteristics implies knowledge of the energy of the DMI. Different methods and approaches are used for the measurement of DMI energy. The first method is Brillouin light spectroscopy which directly determines the effective DMI energy by measuring the frequency difference between negative (Stokes) and positive (anti-Stokes) spin-wave frequencies in Damon-Eshbach geometry [3, 4, 12]. This method is quite reliable, but the measurements are time-consuming. It is worth noting that light falling at non-zero angles to a metal surface only penetrates depths comparable with a skin depth for light in metals (several tenths of nanometers) [13]. The thickness of superlattices may be larger, hence the information given by Brillouin light spectroscopy may only refer to the top layers of the structures. The second method is based on the asymmetrical propagation of chiral Neel DWs under the influence of in-plane (IP) and out-of-plane (OP) magnetic fields [14, 15]. Later, it was shown that the definition of the effective DMI energy by this method in a creep regime is not straightforward, and one needs to take into account the antisymmetric contribution [16, 17] or work in a flow regime with extremely high velocities [18]. This method is suited for systems with large energies of PMA and isolated domains, but it does not work in systems with labyrinth domain structures. There are also various other schemes for measuring DMI: DW propagation via the SOT effect in the presence of IP magnetic fields [19], magnetic droplet nucleation [20], and investigating the microstructure of DWs in combination with micromagnetic modelling [21]. These methods either have insufficient accuracy or aren't applicable in the case of a spin spiral magnetic state.

Under these circumstances, a method based on the comparison of the periodicities of experimentally measured labyrinth structures with micromagnetically simulated ones seems to be straightforward [6, 22]. It only needs knowledge of macroscopic magnetic parameters, which can be measured or found in the literature, and demagnetized domain structures, which may be obtained by various experimental schemes. Moreover, an analytical model for calculating DMI has been introduced [23] and already applied in some papers [24, 25]. However, the aforementioned analytical model did not take into account the possible $z$-dependence of the layers' magnetic structure. Recently, it was shown that dipolar interlayer interaction may lead to the existence of hybrid chiral DWs [26]. The competition between interlayer dipolar and DMI interactions leads to flux-closured DWs configurations which have a completely different energy when compared with the energies of DWs with a fixed internal magnetic structure. Therefore, an accurate model which includes all the layers in the simulation is needed to correctly take into account interlayer dipolar interaction.

The existence of strong DMI was established in symmetric epitaxial $[Co/Pd]_N$ superlattices experimentally and by micromagnetic modeling [27] and in polycrystalline $[Co/Pd]_N$ multilayers by asymmetric DWs propagation [28]. In the present paper we tested the method of comparing the periodicities of experimentally measured labyrinth structures with micromagnetically simulated ones on epitaxial $[Co/Pd(d_{Co})]_5$ symmetric superlattices with Co layers of different thicknesses. The structural and magnetic parameters of $[Co/Pd(d_{Co})]_5$ multilayered structures are investigated. The accuracy of the calculations and the method for the determination of the effective DMI energy are also discussed. Results of the current paper are compared with our previous investigation of DMI in the $[Co/Pd]_N$ superlattices with different numbers of Co/Pd bilayers.

## II. EXPERIMENT

The superlattices were grown in a molecular-beam epitaxy chamber with a base pressure of $3 \times 10^{-10}$ Torr. Si(111) wafers misoriented towards [112] by 0.1° were used as substrates. Wafers were cut into 4 mm × 13 mm pieces, cleaned by acetone, isopropyl and deionized water in an ultrasonic bath, then dried and loaded in the vacuum chamber. After indirect heating at a temperature of 500 °C, the substrates were flash-heated several times by direct current at 1200 °C by and then cooled down to near room temperature. The rates of growth for Cu, Co, and Pd were 0.9, 0.22, and 0.2 nm/min, respectively. The rates of deposition were monitored by a quartz crystal microbalance. The temperature of the substrates was varied from 75 °C during the deposition of the Cu buffer layer, to 110 °C during the deposition of the top Co and Pd layers. Changes in the temperature of the samples during the deposition of different materials were caused by the different radiative heating of the samples from the effusion cells. Epitaxial [Co($d_{Co}$ nm)/Pd(2 nm)]$_5$ superlattices were grown on a Si(111)/Cu(2 nm)/Pd(3 nm) surface. A Cu(2 nm) buffer layer was formed on a Si(111) substrate to prevent the intermixing of Pd and Si and to initiate epitaxial growth of fcc Pd(111). The thickness of the cap Pd layer was 3 nm, which is sufficient to prevent oxidization of the structure. We chose the number of Co/Pd bilayers to be equal to 5 to minimize the influence of dipolar coupling on the inner magnetic structure of the DWs in the superlattices. The thickness of the Co layers $d_{Co}$ was varied between 0.6 and 5 nm. In the following text, a superlattice with a given thickness of Co, for example, 1.2 nm, will be denoted as Co(1.2).

The growth processes and roughness were investigated *in situ* using a scanning tunneling microscope (STM) manufactured by Omicron. The lattice period of the metal layers during growth and their structure were analyzed using reflection high energy electron diffraction (RHEED), made by Staib Instruments. RHEED measurements were done simultaneously with the deposition of the samples. Magnetic characterization of the samples was carried out using a vibrating sample magnetometer (VSM), with magnetic fields up to 27 kOe, manufactured by Lakeshore. The magnetic structure was measured by a magnetic-force microscope (MFM) developed by NT-MDT. MFM images were obtained in the switched-off feedback loop mode using MFM-HM tips manufactured by NT-MDT. The typical distance between the sample surface and magnetic tip was 50 nm.

Micromagnetic simulations were carried out using MUMAX3 software [29]. We used both accurate and single z-cell effective models. All the layers were set explicitly in the accurate model, except in the Co(1.4) samples where the thickness of the Pd layers was taken as 2.1 nm in order to fit the simulation grid. The effective model implies that all the magnetic parameters of each FM layer are averaged in a certain way over the whole FM + NM period [22]. Therefore, one micromagnetic cell corresponding to one FM/NM bilayer, and $N$ cells are needed to model the [FM/NM]$_N$ multilayered structure in the z direction. In this case economy of the time of calculations in not significant, because usually the thickness of a FM layer is less than the thickness of a NM layer, and the number of magnetic cells remains the same as in the case of the accurate model. To go further, one may combine all repetitions and describe the entire [FM/NM]$_N$ structure by one cell in the z-direction. The size of the cell in z-direction has to be size of the cell for one FM/NM period multiplied by the number of FM/NM bilayers. Using of the single z-cell effective model significantly speeds up the calculations. The lateral sizes of the simulation areas were 2 μm × 2 μm for the calculation of the domain structures (4 μm × 4 μm in the case of Co(1) superlattices) and 1 μm × 1 μm for the calculation of the hysteresis loops. The lateral cell size was 2 nm × 2 nm. The cell size in the direction of the normal to the surface was varied to fit the simulation grid defined by thickness of the layers, but never exceeded the thickness of the Co layers. Two-dimensional periodic boundary conditions with 10 repetitions in lateral directions were used in the modeling. The total energy of the system was minimized by the steepest conjugate gradient method, with the built-in function MINIMIZE(). The stopping criterion for energy minimization, MINIMIZERSTOP, was set to $5 \times 10^{-5}$. The saturation magnetization and energies of PMA were determined from the experiment. The considered value of the exchange constant was 25 pJ/m because using this value in the previous paper [27] led to the best coincidence between the simulations and the experiment.

## III. RESULTS AND DISCUSSION

### A. Structural characterization

Epitaxial growth in this system is confirmed by RHEED. RHEED streaks are observed in all samples during the growth of [Co($d_{Co}$)/Pd(2 nm)]$_5$ structures. The RHEED pattern of the Si(111)/Cu(2 nm)/Pd(3 nm)/Co(1 nm) surface is shown in Fig. 1(a), as an example. The structure and growth processes of the crystalline [Co($d_{Co}$)/Pd(2 nm)]$_5$ superlattices are similar to those for [Co/Pd(111)]$_N$ superlattices, which were thoroughly described in our previous paper [27]. In this section we focus on strain relaxation in the Co layers which depends on the Co thickness. The bulk lattice parameters of fcc-Pd and fcc-Co are 0.389 and 0.355 nm, respectively. Hence, Co is largely (9.6%) strained when deposited on the Pd surface. The growth of Co on the Pd(111) surface is incoherent from the beginning and is accompanied by the incorporation of misfit dislocations [30]. The strains in Co gradually relax but the degree of strains in the top Co atomic layers depends on the thickness of the Co that has already been grown. The thickness dependencies of the lattice parameters of the first three material layers, except for Cu buffer layer in the [Co($d_{Co}$)/Pd(2 nm)]$_5$ samples, are shown in Fig. 1(b). The parts of the curves where the lattice parameter decreases correspond to the Co growth. The lattice parameter restores to the Pd bulk value after the deposition of a Pd layer on top of a Co layer, if the thickness of the Co layer is 1.6 nm and less. If the first Co layer is thicker, then 2 nm of deposited Pd is not enough to restore the lattice parameter to the Pd bulk value, like it is in the case of the Co(2.4) superlattice in Fig 1(b). Strains in the bottom Pd/Co interfaces of the second and upper Co layers decrease and begin to depend on the Co thickness, if it is thicker than 1.6 nm. We focused on the Co thickness interval less than 1.6 nm because the effective PMA is positive in this thickness interval. Distribution of the strains in all the layers in the Co(0.8) and Co(1.6) superlattices is indicated in Fig. 1(c). It is evident that strains in the top Co/Pd interface slightly decrease in the topmost layers but may be described by the averaged value. The dependencies of the strains in the bottom Pd/Co and top Co/Pd interfaces on the Co thickness are shown in Fig. 1(d). Asymmetry of strains between the bottom and top interface Co layers increases with increasing Co thickness.

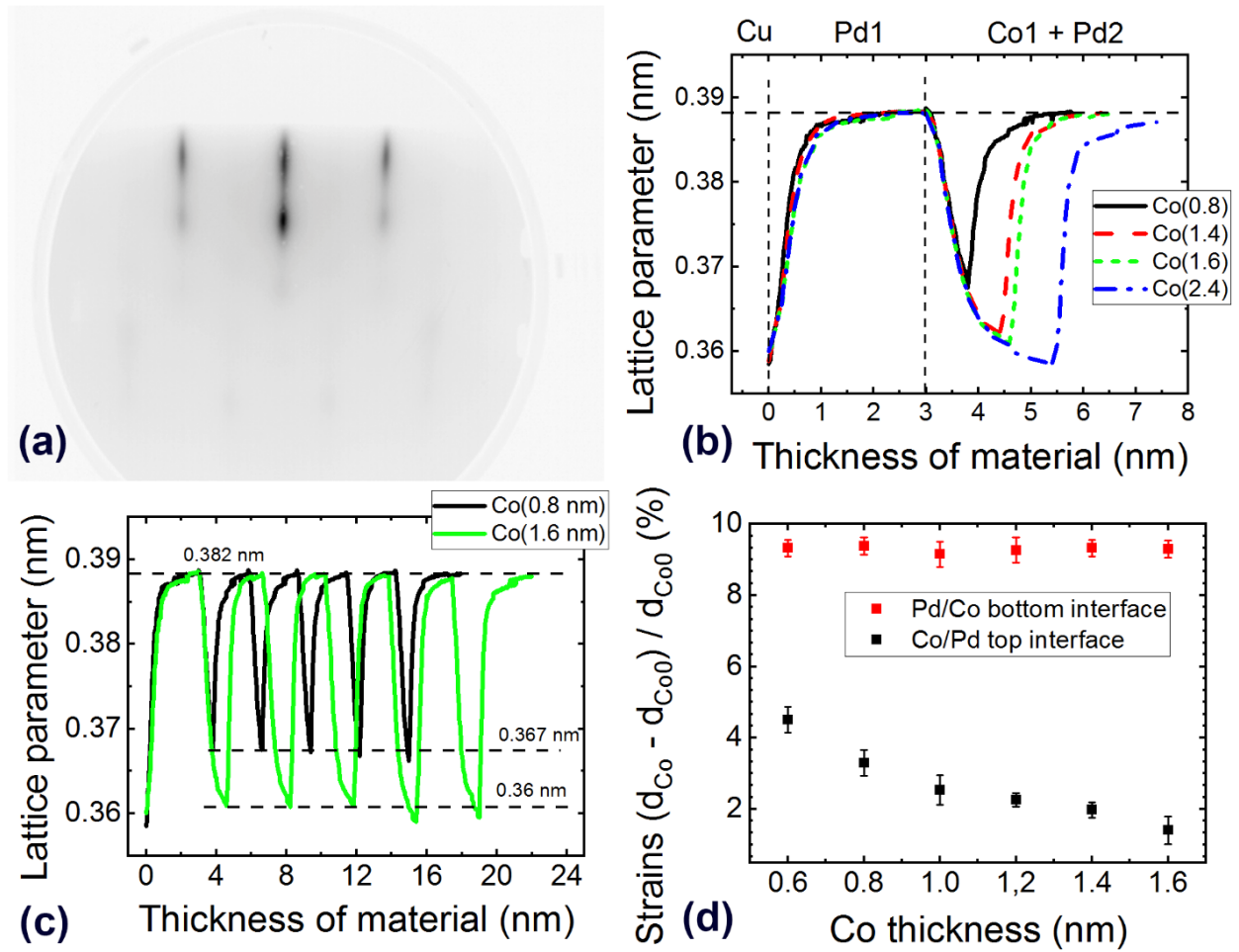

FIG. 1. (a) RHEED pattern of the Si(111)/Cu(2 nm)/Pd(3 nm)/Co(1 nm) surface. (b) Lattice parameter in the first three material layers of the different [Co($d_{Co}$)/Pd(2 nm)]$_5$ structures. (c) Distribution of the lattice parameters in the Co(0.8) and Co(1.6) superlattices. (d) Strains in the bottom Pd/Co and top Co/Pd interfaces, as functions of the Co thickness, $d_{Co0}$ is the bulk lattice parameter of fcc-Co.

It should be noted that we did not detect any meaningful differences in the roughness on the top of the [Co($d_{Co}$)/Pd(2 nm)]$_5$ structures in the Co thickness interval of 0.6 to 1.6 nm. This may be related to the fact that Co tends to smooth the surface of the underlying Pd layer in the initial stages of the growth [31]. Hence, if the thickness of the Co layers is not quite large enough then the growth of Co on Pd does not lead to an increase in the roughness.

### B. Magnetic properties and magnetization reversal

Magnetization reversal processes in the [Co($d_{Co}$)/Pd(2 nm)]$_5$ multilayers strongly depend on the Co thickness. The OP and IP hysteresis loops of the samples with a $d_{Co}$ less than 1.6 nm are outlined in Fig. 2 (a) and (b), respectively. The PMA decreases with an increase of the Co thickness, hence saturation fields in the IP hysteresis loops decrease. An increase in the Co thickness also leads to an increase of the magnetostatic energy of the system. These two factors result in the self-demagnetization of superlattices with thick Co layers. Magnetization reversal processes in the OP directed magnetic fields change with increasing Co thickness. At low Co thickness ($d_{Co} < 0.6$ nm), when the energy of PMA is very large, the hysteresis loops are rectangular and magnetization reversal occurs by the switching of the magnetization in the entire sample. An increase in $d_{Co}$ leads to the emergence of "tails" in the OP hysteresis loops in the regions of loops near saturation. These "tails" are related to the existence of domains which are quite stable in large magnetic fields and needed additional energy to be removed. These "tails" extend with increasing Co thickness ($d_{Co} = 0.6 - 0.8$ nm), however, normalized remanent magnetization still remains equal to unity. This particular thickness interval is interesting for practical applications because isolated domains stabilized in large negative magnetic fields remain almost unchanged after switching off the magnetic field. Further increases in the Co thickness make it possible to nucleate negatively magnetized domains, even in positive magnetic fields. Nucleated domains develop into a labyrinth domain structure in the remanent state. Negatively magnetized domains grow in size in negative magnetic fields, the labyrinth structure transforms into isolated stripe domains, then into bubble domains in the fields near negative saturation ($d_{Co} = 1 - 1.6$ nm). In this Co thickness interval the isolated magnetic domains develop to the labyrinth state in the remanent state, which is not favorable for racetrack memory devices. A more detailed description of the magnetization reversal processes of self-demagnetizing superlattices may be found in the paper of J.E. Davies et al. [32].

The dependence of the magnetic moment, normalized to the unity of an area, on the Co thickness measured in the [Co($d_{Co}$)/Pd(2 nm)]$_5$ multilayers is shown in Fig. 2(c). It is well known that Pd interface layers are magnetically polarized in the vicinity of the Co layers [33]. Therefore, the net magnetic moment per unity of an area may be described as

$$\frac{m}{S} = 5M_{s, Co}d_{Co} + 10M_{s, Pd}d_{Pd\,pol}, \qquad (1)$$

where $m$ is a magnetic moment of the sample, $S$ is an area of the film, $M_{s, Co}$, and $M_{s, Pd}$ are the saturation magnetizations of Co and polarized Pd, respectively; $d_{Co}$, and $d_{Pd\,pol}$ are the thicknesses of the deposited Co layers, and the polarized interface Pd layers, respectively. The formula takes into account five Co layers and ten polarized

Pd interface layers. Interception of the linear fitting of $m(t_{Co})/S$ dependence with the y axis gives a positive sum magnetic moment per unit area, $10M_{s,\,Pd}d_{Pd\,pol} = 7.5 \times 10^{-4}$ A, induced in all the Pd interface layers. Using the value of the saturation magnetization of Pd, $M_{s,\,Pd} = 0.31 \times 10^6$ A/m [34], the effective thickness of each of the polarized Pd interface layers is calculated as $d_{Pd\,pol} = 0.24$ nm, which agrees well with the result from our previous paper, 0.2 nm. The latter value is used in this paper. The slope of the $m(d_{Co})/S$ dependence in the [Co($d_{Co}$)/Pd(2 nm)]$_5$ multilayers gives a value of saturation magnetization of $M_{s\,Co} = 1.43 \times 10^6$ A/m, which is close to the bulk value for Co. If one neglects the Pd polarized layers with low magnetization and assumes that all the magnetic material is concentrated in the Co layer, then the value of the saturation magnetization in the [Co($d_{Co}$)/Pd(2 nm)]$_5$ superlattices is a function of the Co thickness:

$$M_s = \frac{M_{s,Co}d_{Co} + 2M_{s,Pd}d_{Pd\,pol}}{d_{Co}}. \qquad (2)$$

These values of saturation magnetization were used in the calculation of the effective energy of PMA, $K_{eff}$. All the magnetic parameters are listed in Table I. The dependence of the $K_{eff} \times d_{Co}(d_{Co})$ is outlined in Fig. 2(d). Using the approach proposed in our previous paper [31], we calculated the magnetoelastic and interface contributions to the surface energy of PMA in this system. Fitting the experimental data gives the interface magnetic anisotropy contribution $K_s = 1.1$ mJ/m² and magnetoelastic surface and volume contribution, $K_{s,\,MEA} = 0.81$ mJ/m² and $K_{v,\,MEA} = 0.16$ MJ/m³, respectively. These values agree well with our results reported for single-layered Pd(2.25 nm)/Co($d_{Co}$)/Pd(2.25 nm) films.

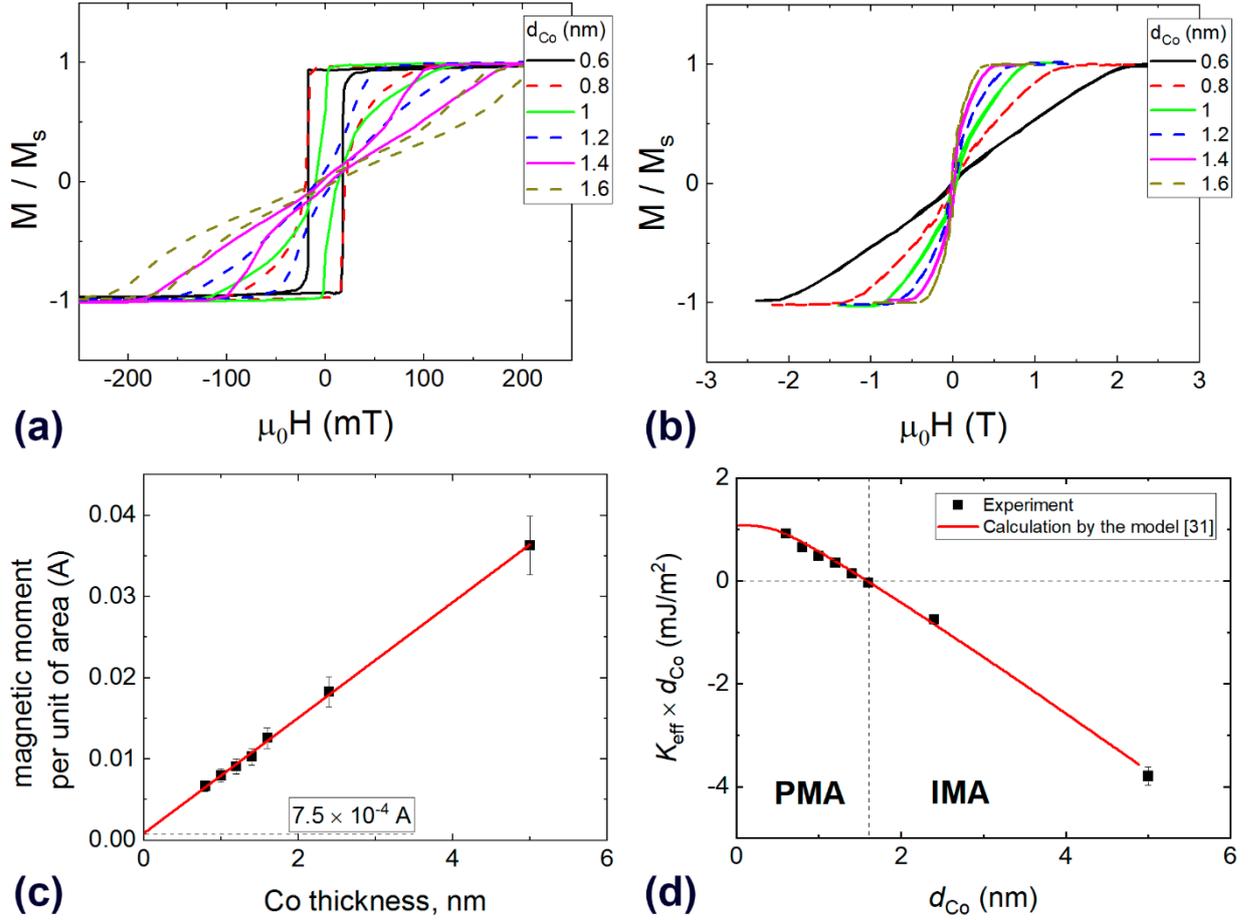

FIG. 2. (a) OP and (b) IP hysteresis loops of crystalline $[Co(d_{Co})/Pd(2\ nm)]_5$ superlattices. The dependencies of (c) magnetic moment per unit of area, and (d) $K_{eff} \times d_{Co}$ on the Co thickness in the $[Co(d_{Co})/Pd(2\ nm)]_5$ superlattices.

### C. MFM measurements of the demagnetized domain structures

The periodicity of the magnetic structures was determined by statistical analysis of the 20–30 random profiles between adjacent domains. The periodicity and anisotropy of the demagnetized labyrinth domain structure depends on the direction of an alternating demagnetizing field with a decreasing amplitude. If the magnetic field is oriented OP, then the magnetic structure is isotropic. If the magnetic field is oriented IP, then magnetic stripes orient towards the field axis and the periodicity is less than in the previous case. The OP and IP demagnetized domain structures of $[Co(d_{Co})/Pd(2\ nm)]_5$ superlattices are shown in the Fig. 3 (a) and (b), respectively.

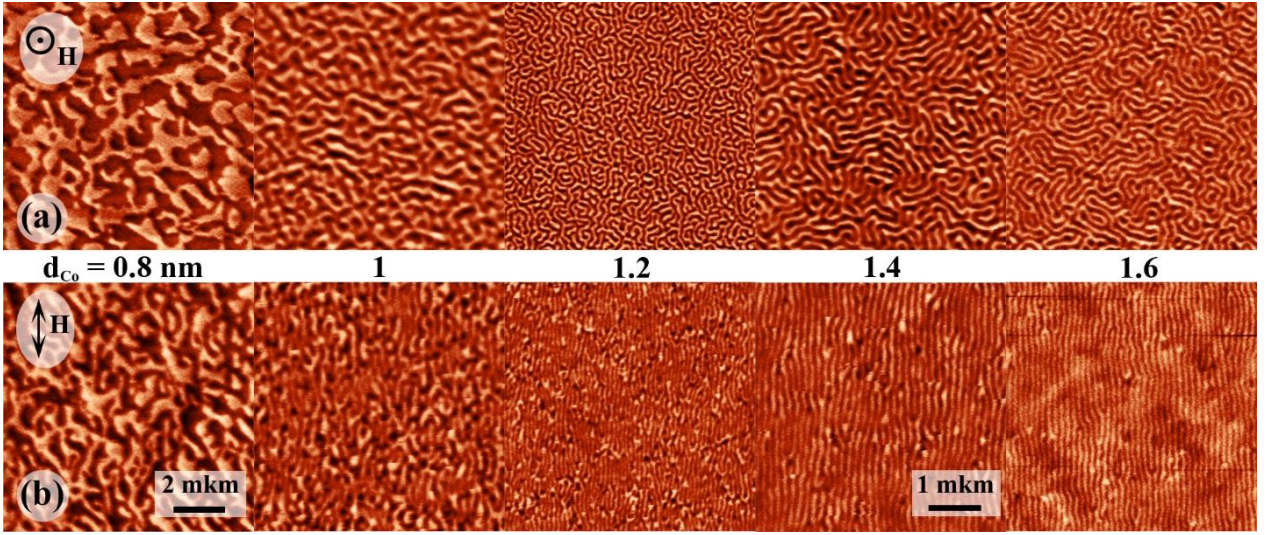

FIG. 3. MFM images of (a) OP and (b) IP demagnetized domain structures of the $[Co(d_{Co})/Pd(2~nm)]_5$ superlattices. Note that scale is halved reduced in the case of $d_{Co} = 1.4$ and $1.6$ nm.

Regular domain patterns are observed when the Co thickness in the $[Co(d_{Co})/Pd(2~nm)]_5$ superlattices is 1 nm and larger. The size of the stripe domains decreases with an increase of the Co thickness (Fig. 4). This phenomenon may be explained solely by the decreasing energy of PMA and the increasing magnetostatic energy of the system. Similar domain structures were observed in systems in which the DMI was supposed to be absent [35]. However, an addition of DMI in such systems lowers the energy of DWs [23] and hence, leads to a decrease of domain sizes when compared with the case when the energy of effective DMI is zero.

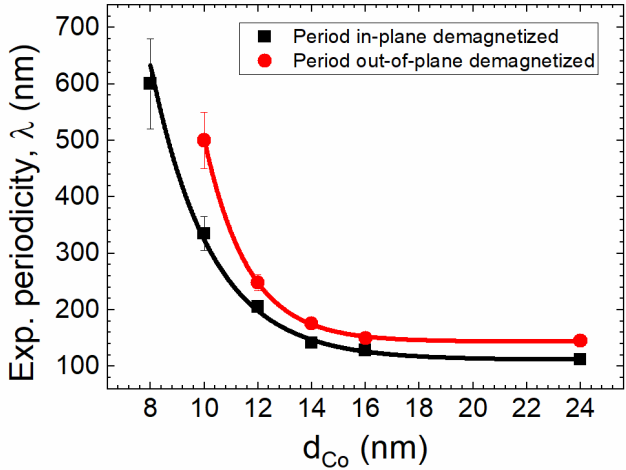

FIG. 4. Periodicities of the IP and OP demagnetized stripe domain structures as functions of the Co thickness in the $[Co(d_{Co})/Pd(2~nm)]_5$ superlattices.

### D. Micromagnetic simulations

Micromagnetic simulations were carried out to estimate the energy of effective DMI in this system. Experimentally measured periodicities of the labyrinth domains are compared with periodicities of simulated magnetic structures with different effective DMI energies. Since the periodicity of experimental domain structures depends on the direction of demagnetizing field, two effective energies of DMI may be obtained: the first one if the simulated structures are compared with IP demagnetized images, and the second one, if the simulated structures are compared with OP demagnetized images. They will be referred to in the text as IP and OP effective DMI energies ($D_{\text{eff IP}}$ and $D_{\text{eff OP}}$), respectively. In most of the papers in which the energy of the DMI is evaluated by the analysis of the periodicity of the stripe domain structures, the authors use either IP or OP demagnetized domain structures and do not show comparisons of the results obtained using both of these structures. IP demagnetized structures were used in Ref. [26, 36], and OP demagnetized structures or magnetic structures in the remanent state after OP saturation were used in Ref. [6, 37−39].

In micromagnetic simulations, we used a monodomain magnetic structure with two skyrmions as an initial state. The diameter of the skyrmions was 100 nm. The magnetic configuration was calculated in the remanent state by the evolution of isolated skyrmions into the labyrinth domain state. In this case, the simulated magnetic structures are isotropic and match the OP demagnetized experimental images well. Therefore, it is natural to compare the OP demagnetized structures with the simulated ones in the present case. In most of the papers a random initial state is used for the micromagnetic modeling of the labyrinth domain structures [7, 23, 40]. However, the random magnetic state is very nonequilibrium and is not possible in real structures. Relaxation from the random state in zero magnetic field leads to the presence of a large number of skyrmions in relaxed states, which does not coincide with the experimental images. Conversely, an experiment indicated that the stripe domain structure develops from single isolated skyrmions when magnetization reversal occurs in the OP magnetic fields [32].

The dependencies of the periods of simulated domain structures on the effective energy of the DMI for $d_{\text{Co}}$ = 1, 1.2, 1.4, and 1.6 nm in the [Co($d_{\text{Co}}$)/Pd(2 nm)]$_5$ superlattices are shown in Fig. 5(a-d), respectively. A comparison of the results obtained using the accurate and single z-cell effective models is also shown. We noticed a growing discrepancy between the results obtained by singe z-cell effective and accurate modelling with an increase of the Co thickness. Moreover, the results obtained by singe z-cell and N-z-cell models are quite similar. We relate this to the growing influence of magnetostatic energy in the case of thicker Co films. Effective models do not correctly take into account the magnetostatic interaction between the layers. Dipolar interaction may strongly influence both the size of the domains and internal DWs structure. It was shown that dipolar interaction leads to hybrid chiral DWs with a Bloch-like structure in the center, even in the systems with non-zero DMI [26,6 35]. But when DMI become stronger, the Bloch regions are pushed out to the surface of the films and the DWs become homochiral. The domain structure of the homochiral DWs becomes uniform throughout all the layers of the structure. Therefore, the results of the effective and accurate modeling converge to a single curve at large DMI energies. Based on the aforementioned results we claim that the accurate model has to be used in the micromagnetic modeling, or a careful comparison of the results obtained by effective and accurate models needs to be carried out. One possible way of taking into account the dipolar coupling in the effective model may be the insertion of a block of RKKY coupling in MUMAX3, with zero antiferromagnetic coupling between the effective layers. Using this trick, the exchange interaction between the effective layers is switched off, while the magnetostatic energy is correctly calculated. However, this approach needs to be thoroughly verified. The IP and OP effective DMI energies of the samples with $d_{Co}$ = 0.8–1.6 nm, using the results from the micromagnetic simulations based on the accurate model, are listed in Table II.

TABLE I. Parameters of the epitaxial [Co($d_{Co}$)/Pd(2 nm)]$_5$ superlattices.

| $d_{Co}$ (nm) | $A$ (pJ/m) | $M_s$ (kA/m$^3$) | $\mu_0 H_{eff}$ (T) | $K_{eff}$ MJ/m$^3$ | $\lambda_{OP}$ (nm) | $\lambda_{IP}$ (nm) |
|---|---|---|---|---|---|---|
| 0.6 |  | 1626 | 1.89 | 1.5 | - | - |
| 0.8 |  | 1575 | 0.84 | 0.66 | - | 600 ± 50 |
| 1 | 25 | 1544 | 0.63 | 0.49 | 500 ± 40 | 335 ± 25 |
| 1.2 |  | 1523 | 0.39 | 0.29 | 248 ± 14 | 205 ± 10 |
| 1.4 |  | 1508 | 0.14 | 0.1 | 176 ± 8 | 141 ± 5 |
| 1.6 |  | 1497 | -0.03 | –0.02 | 150 ± 7 | 128 ± 3 |

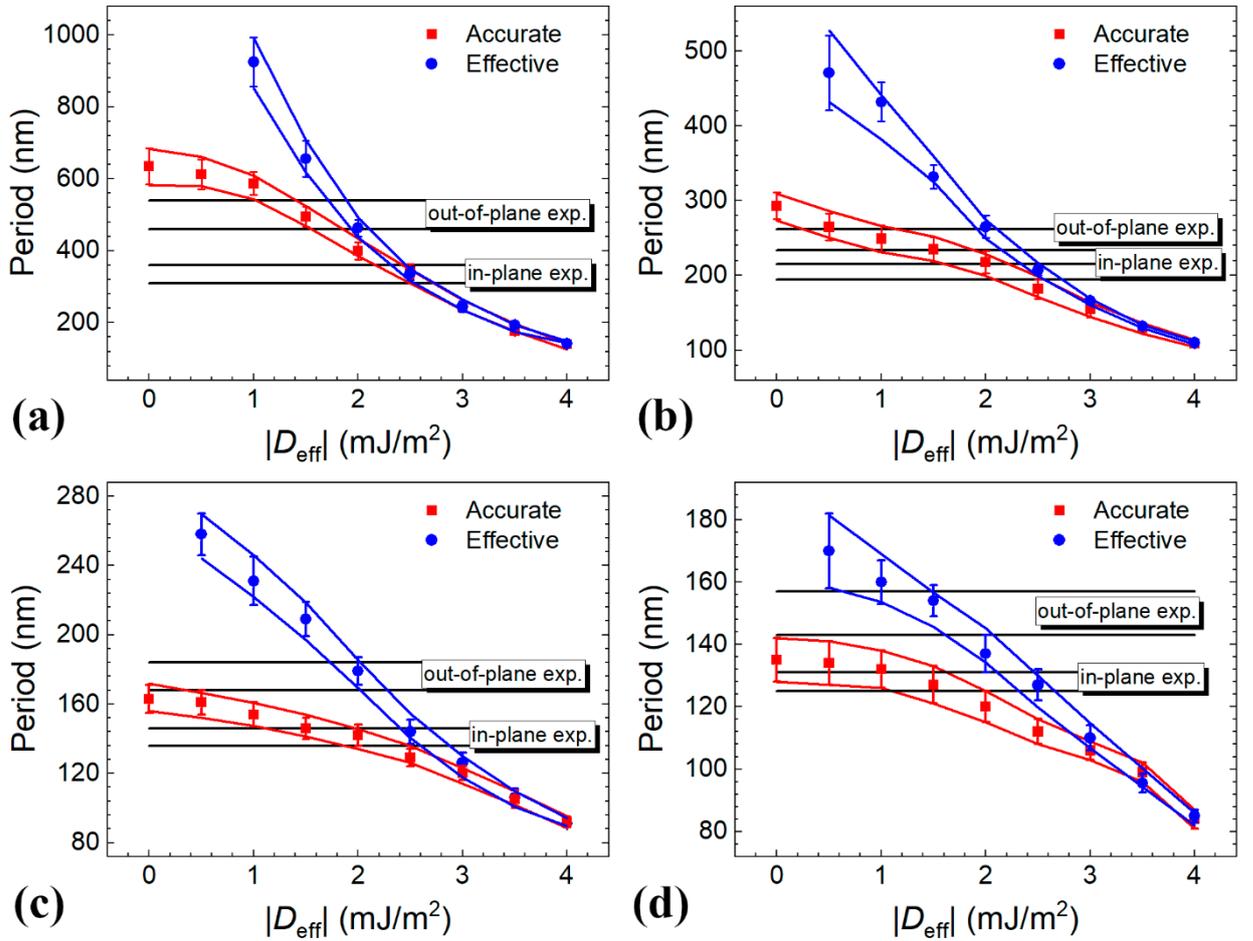

FIG. 5. Comparison of the simulated periods of labyrinth domain structures with the experimentally obtained ones for the $[Co(d_{Co})/Pd(2\ nm)]_5$ superlattices with $d_{Co}$ equal to (a) 1, (b) 1.2, (c), 1.4, and (d) 1.6 nm. Solid lines represent possible smoothed lower and upper values of the determined periods, taking into account the errors of the measurements.

The effective energy of Co(0.8) superlattices could not be deduced by the above approach because skyrmions do not transform into the labyrinth domain state in the simulations in zero magnetic fields using the magnetic parameters of these samples. This is expected, since the normalized remanent magnetization $M_r/M_s$ is unity in the OP hysteresis loops of Co(0.8) structures. Therefore, to deduce the effective DMI energy in the Co(0.8) superlattices we tried to compare the experimentally measured hysteresis loops of Co(0.8) superlattices with the simulated ones with different effective DMI energies. This approach is not very accurate, because the simulated hysteresis loops of Co(0.8) samples weakly depend on the effective energy of DMI. We found that the possible effective DMI energy values in the Co(0.8) superlattices were equal to $1 \pm 0.6\ mJ/m^2$. In the previous paper, we investigated the dependence of the effective DMI energy on the number of Co/Pd bilayers ($N = 1-20$) in the $[Co(0.8\ nm)/Pd(2\ nm)]_N$ superlattices [27]. The present series of samples intersects with the past series at $d_{Co} = 0.8$ nm. Taking into account the results of the previous paper, more definite effective energies of DMI of Co(0.8) superlattices, $|D_{eff\ OP}|$ and $|D_{eff\ IP}|$ were determined as $1.2 \pm 0.5$ and $1.6 \pm 0.4$ mJ/m$^2$, respectively. It is worth mentioning that the $|D_{eff\ IP}|$ values established on the results of the previous work are slightly larger than the possible energy values determined by the hysteresis loops analysis. See Appendix A for more details regarding the method of the estimation of the energies of effective DMI in the Co(0.8) samples.

### E. Discussion

A net surface DMI energy $D_s$, may be derived using the value of effective DMI energy using the following formula:
$$D_S = D_{eff} d_{Co}. \tag{3}$$
The effective and surface DMI energies estimated by the comparison of the periodicities of magnetic structures obtained by accurate modeling and MFM are outlined in Fig. 6 (a) and (b), respectively. The error values are determined using the uttermost points of the intersections of the experimental and simulated periods in Fig. 5. In a simple phenomenological model, taking into account that DMI is of an interface origin in this system, the surface DMI energy is considered to be a constant. Thus, the dependencies of the IP effective and surface DMI energies on the Co thickness are more reasonable than for the respective OP energies. The $|D_{S\ IP}|$ curve more or less satisfies the phenomenological model. The $|D_{S\ IP}|$ values are nearly constant if $d_{Co}$ is more than 1 nm, but they significantly decrease in lower Co thickness intervals, which goes against the phenomenological model. In spite of the relatively large error in the determination of DMI energies in the Co(0.8) superlattices, the upper bound of $|D_S| = 1.6$ pJ/m is strictly determined by monitoring the self-demagnetization of the system in the remanent state and comparing this value with the previous results (see Appendix A). However, a decrease of the surface DMI energy in the low FM layer thickness regime is frequently observed in various systems [4, 41, 42]. This may be explained by structural inhomogeneity or by the degradation of the top FM/NM interface.

The dependence of $|D_{S\ OP}|$ on the thickness of Co is at maximum in the Co thickness interval of 1−1.2 nm, with an abrupt decrease in the large Co thicknesses regime, which could not be explained by the phenomenological model. The only explanation of this is that the net surface DMI energy is not a constant but it depends on the Co thickness. Indeed, while the strains in the bottom Pd/Co interfaces are the same for all of the samples, the top Co/Pd interfaces undergo different strains. Strains in the top Co/Pd interfaces decrease with increasing Co thickness. If we suppose that non-zero net DMI in this system is proportional to the asymmetry of the strains between the bottom Pd/Co and top Co/Pd interfaces, then the net surface DMI should increase with an increase of the Co thickness, which is not observed in both the cases of IP and especially OP DMI energies. Therefore, to explain the behavior of $|D_{S\ OP}|$ and its dependence of Co thickness we may suppose that that the dependence of the net surface DMI energy on the strains in the top Co/Pd interfaces is not simply linear but reaches a maximum in the Co thickness interval of 1–1.2 nm.

Another point that needs discussing is the difference between the absolute values of DMI energies estimated by comparing the simulated images with different experimentally obtained MFM images of magnetic structures. IP effective DMI energies in the investigated superlattices are rather large when compared with reported data for other systems with the same thickness of Co layers, as indicated in Table II. The values for the IP effective DMI energies at large Co thicknesses near the spin reorientation transition of the magnetization to the plane of the structures are especially doubtful. The values for the OP effective DMI energies seem much more reasonable, taking into account that the investigated system is symmetric by the composition. Since there is a strong dependence of the experimentally obtained periodicity of the domains on the direction of the demagnetizing field, then it may be revealed in micromagnetic simulations. We also tried to obtain IP demagnetized structures by simulating domain patterns going from IP saturation to a remanent state. The details are in Appendix B. Indeed, we found that the periodicities of the aligned IP demagnetized structures are less than the periodicities of OP demagnetized ones. However, convergence of the simulations was very slow.

TABLE II. Effective DMI energies obtained in this paper and in other systems with the same Co thickness,

| $d_{Co}$ (nm) | $|D_{eff\ IP}|$ (mJ/m$^2$) | $|D_{eff\ OP}|$ | System | $|D_{eff}|$ (mJ/m$^2$) |
|---|---|---|---|---|
| 0.8 | 1.6 ± 0.4 | 1.2 ± 0.5 | [Co/Ir(1 nm)/Pt(1 nm)]$_5$ [26] | 1.64 |
| 1 | 2.4 ± 0.25 | 1.4 ± 0.4 | Pt(2 nm)/Co/IrMn(2.4) [43] | 1.22 |
| 1.2 | 2.1 ± 0.5 | 1.1 ± 0.6 | Pt/Co/AlO$_x$ [4] | 1.15 |
| 1.4 | 1.8 ± 0.6 | 0 ± 0.5 | Pt/Co/AlO$_x$ [44] | ≅ 0.46 |
| 1.6 | 1.4 ± 0.6 | 0 | Pt/Co/AlO$_x$ [4] | 0.7 |

If we suppose that the OP effective DMI energies are correct, then one may notice that the $\lambda(D_{eff})$ dependency simulated for the Co(1.6) sample reaches a value that is lower than that for a band of experimentally obtained OP demagnetized periods in FIG. 5(d). A possible explanation for this discrepancy may be related to the accuracy of determining the energy of PMA. The experimental domain structures were measured in the center of the samples, where the magnetic properties are the most uniform. However, hysteresis loops, with a help of which energy of PMA was calculated, were measured from entire sample. We noticed a slight reduction in PMA at the edges of the samples. This may be due to lower temperatures at the edges during the annealing of the Si(111) substrate or the migration of the metal atoms from the sample plates during annealing and the formation of islands of silicide. This effect is not significant, but it may slightly decrease the PMA averaged over the whole sample when compared with the PMA in the central areas. Substitution of the slightly larger energies of PMA in the simulations will shift the $\lambda(D_{eff})$ curves towards larger values of $\lambda$. Another possible explanation may be related to the appropriate choice of exchange constant. In the previous paper, we showed that an increase in the exchange constant leads to an upward shift of the $\lambda(D_{eff})$ curves, which solves the problem [27]. It is worth noting that playing with micromagnetic parameters and shifting $\lambda(D_{eff})$ curves up to fit the $\lambda(D_{eff})$ dependency of Co(1.6) superlattice to experimental values will lead to an increase of $D_{eff\,IP}$ in all the samples and $D_{eff\,OP}$ determined from the Co(1), Co(1.2), and Co(1.4) samples.

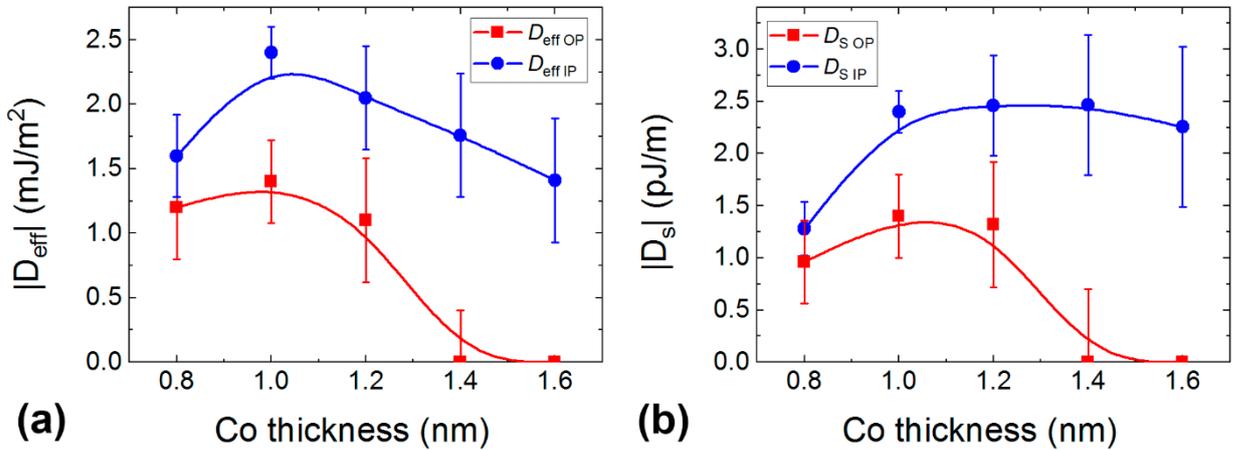

FIG. 6. The absolute values of (a) effective, and (b) surface DMI energies depending on the Co thickness in the crystalline $[Co(d_{Co})/Pd(2\,nm)]_5$ superlattices. Solid lines are B-spline interpolation curves.

The results of the present micromagnetic simulations are also a sign of the fact that DMI does not play a decisive role in the demagnetization of the superlattices with low effective PMA. Stripe domains with a relatively small width of approximately 65 nm may be stabilized in the absence of DMI in the Co(1.6) superlattices. Addition of DMI in the simulation does not significantly change the situation if the $|D_{eff}|$ is less than 1.5 mJ/m$^2$, which is a rather large value for structures with such thick Co layers. The errors of definition for the effective DMI energies increase with increasing thicknesses of FM layers towards the thickness of spin reorientation transition, because $\lambda(D_{eff})$ curves intersect the horizontal bar of experimental periods at sliding angles. In this case, it is dipolar coupling, rather than the DMI, which defines the periodicities of the magnetic structures. Micromagnetic modeling shows that the DWs between the stripe domains in the Co(1.6) superlattices with $D_{eff} = 0$ are hybrid [26] with a Bloch-like structure in the central layer and oppositely magnetized Neel caps in the bottom and top layers [35, 45, 46]. Setting $|D_{eff}| = 1.5$ mJ/m$^2$ in this case leads only to changing the magnetic structure of the DW in the central layer to the Neel type, while the magnetic structure of the DWs in the bottommost and topmost layers remains almost the same. It is worth noting that DWs even in the Co(1) superlattices with the largest possible DMI energy of $|D_{eff\ IP}| = 2.5$ mJ/m$^2$ do not possess uniform chirality in all the layers, but demonstrate opposite chirality in the fifth layer to the curling of the magnetization than is favored by DMI and observed in four bottom layers.

In the present paper, we do not focus on the sign of the evaluated effective DMI energies, because the periodicities of the labyrinth domain structures obtained by micromagnetic modeling do not depend on the sign of $D_{eff}$. Based on the behavior of the domains under the influence of IP and OP magnetic fields [14, 15] we may conclude that the DMI induces predominantly right-handed chirality in the Neel DWs in the [Co(0.6 nm)/Pd(2 nm)]$_5$ superlattices. If we suppose that chirality does not change with increasing Co thickness in the [Co($d_{Co}$)/Pd(2 nm)]$_5$ superlattices and use the DMI energy expression as in [29], a *negative* sign of the effective energies derived in the [Co($d_{Co}$)/Pd(2 nm)]$_5$ superlattices should be considered.

### IV. CONCLUSIONS

The structural and magnetic properties of epitaxial [Co($d_{Co}$)/Pd(2 nm)]$_5$ superlattices were investigated with regard to their dependence of the Co thickness. The asymmetry of the strains between the lower Pd/Co and upper Co/Pd interfaces increases with an increase in the Co thickness. The dependence of the energy of PMA on the Co thickness is well fitted by the previously proposed model, which indicates that PMA is due to the interface and magnetoelastic contributions in this system. IP and OP demagnetized structures were experimentally measured in epitaxial [Co($d_{Co}$)/Pd(2 nm)]$_5$ superlattices. The effective and surface energies of the DMI in this system were determined by a comparison of the periodicities of simulated domain structures with the stripe domain structures experimentally measured after the demagnetization of the samples in IP and OP magnetic fields. A comparison of the obtained results with data from the literature indicate that values of OP DMI effective and surface energies are more reasonable than IP DMI energies. Maximal energy of effective DMI is observed in the [Co(1 nm)/Pd(2 nm)]$_5$ superlattices. The appropriate choice of Co thickness provides the opportunity to tune the energy of effective DMI and control the magnetic structures in the epitaxial [Co($d_{Co}$)/Pd(2 nm)]$_5$ superlattices. Taking into account the opportunity of controlling the energy of effective DMI in this system by varying the number of Co/Pd bilayers it may be concluded that the epitaxial [Co/Pd]$_N$ multilayered structure is a system with flexible magnetic parameters, which is promising as a medium for skyrmion memory engineering.

## ACKNOWLEDGEMENTS

The reported study was partially funded by RFBR under the research projects № 18-02-00205 and 18-32-20057 and by the Grant program of the Russian President (MK-5021.2018.2), and the Russian Ministry of Science and Higher Education under the state tasks (3.5178.2017/8.9).

## APPENDIX A: METHOD FOR THE ESTIMATION OF THE ENERGIES OF EFFECTIVE DZYALOSHINSKII-MORIYA INTERACTION IN Co(0.8) SAMPLES.

The approach for estimating the effective energy of DMI by comparing experimental and modeled hysteresis loops is not reliable in the case of Co(0.8) superlattices. Hysteresis loops simulated with $D_{eff} = 0$–$1.8$ mJ/m$^2$ are shown in Fig. 7. As the criteria of coincidence, we analyzed the tilting of the hysteresis loops in the near-zero magnetic fields and the values of the normalized remanent magnetization. The coercive force of the experimental and simulated loops may be slightly different due to structural defects that are not taken into account in the simulations. The hysteresis loops simulated with $D_{eff} = 0.3$ and $1.7$ mJ/m$^2$ definitely do not fit with experimental one. The simulated hysteresis loop with $D_{eff} = 0.3$ mJ/m$^2$ is more rectangular than experimental one, and the normalized remanent magnetization of the hysteresis loop in the case of $D_{eff} = 1.7$ mJ/m$^2$ is less than 1. However, the effective DMI energies that are in the fairly broad range of 0.4 to 1.6 mJ/m$^2$ give identical hysteresis loops in the simulations, which fit the experimental ones quite well.

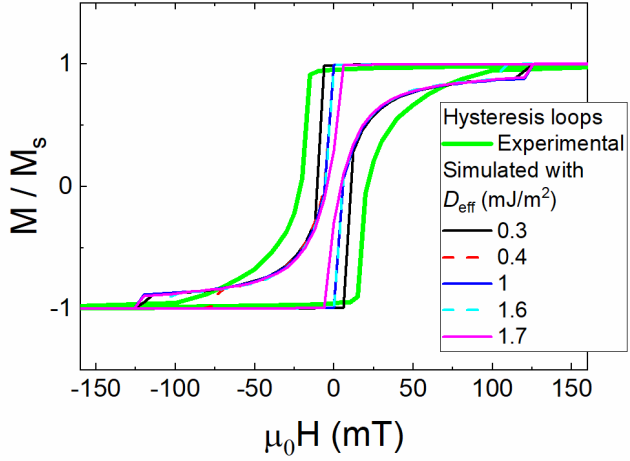

FIG. 7. Experimental and simulated magnetic hysteresis loops of Co(0.8) superlattices.

To increase the accuracy of the measurements we decided to compare the results of our previous investigation [27] with the current results. In the previous paper, we investigated the dependence of the effective DMI energy on the number of Co/Pd bilayers $N = 1-20$ in the [Co(0.8 nm)/Pd(2 nm)]$_N$ superlattices using the single z-cell effective model in micromagnetic simulations. Present series of the samples intersects with $N$-series at $d_{Co} = 0.8$ nm ($N5$ sample in the $N$-series). Hence, the effective DMI energies of Co(0.8) and N5 samples must be the same. We reran the micromagnetic simulations of the labyrinth domain structures for $N$-series using the accurate model instead of the effective. Also, we used the latest value of $D_{eff} = 0.6 \pm 0.1$ mJ/m$^2$ for the N1 samples. We fitted the $D_{eff}(N)$ plots using the exponential function $y = y_0 + Ae^{R_0 x}$, and found the values of $D_{eff\,IP} = 1.6 \pm 0.5$ mJ/m$^2$ and $D_{eff\,OP} = 1.2 \pm 0.5$ mJ/m$^2$ for N5 or Co(0.8) samples, which agree quite well with the results of the hysteresis loops analysis (Fig. 8). It is worth noting that using the accurate instead of the single z-cell effective model does not significantly change the results in the case of $N$-series. This is explained by the fact that the Co layers are rather thin and sufficiently far apart. Hence, interlayer dipolar coupling is less significant in the case of the $N$-series when compared with the Co-series of samples.

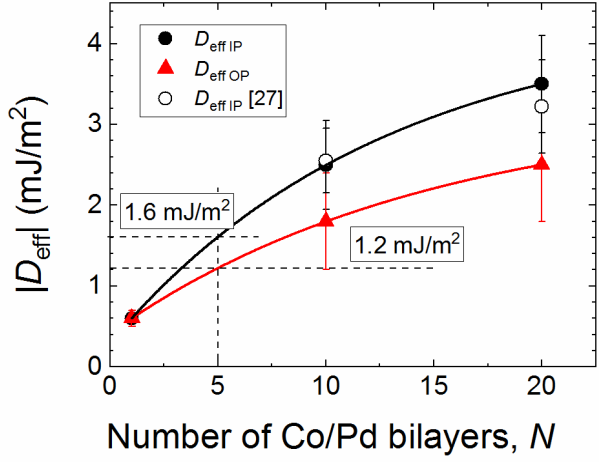

FIG. 8. The absolute values of effective DMI energies depending on the number of Co/Pd bilayers in the crystalline [Co(0.8 nm)/Pd(2 nm)]$_N$ superlattices. Solid lines denote exponential fitting of the obtained $D_{eff}(N)$ dependencies.

## APPENDIX B: MODELING OF IN-PLANE DEMAGNETIZED STRUCTURES

Analysis of the obtained results indicates that magnetic structures simulated from an initial state of several skyrmions in a zero magnetic field are better when compared with OP demagnetized experimental structures. Therefore, we tried to model the IP demagnetization process to compare the results with IP demagnetized experimental structures. The Co(1) superlattice was considered to be the simplest in terms of geometry. The effective DMI energy was equal to 1.5 mJ/m². Firstly, we used uniform magnetization in the x-direction as an initial state and simulated magnetic structures going from positive magnetic fields of 1 T oriented towards the x-direction to the remanent state. Two-dimensional periodic boundary conditions in lateral directions were used in the modeling. In this case the anisotropic stripe domain structure with the period of 118 nm nucleated in $\mu_0 H$ = 198 mT and was not changed anymore (Fig. 9(a)). Obviously, this periodicity is too small and energetically unfavorable in the case of Co(1) superlattice. We suppose that the periodic boundary conditions prevent the annihilation of redundant DWs in the case of an ideally symmetric structure. Therefore, in the initial state we added two magnetically frozen skyrmions as disturbing centers. In this case, the stripe domain structure nucleated already in $\mu_0 H$ = 700 mT and changed its periodicity during the magnetization reversal. The average periodicity in the remanent state was 380 nm, which is a reasonable value (Fig. 9(b)). However, the calculations took several days for one magnetic structure obtained at a certain DMI energy, which is too long. A key factor in stabilizing uniaxial anisotropic stripe domain state is the gradual decrease of the IP magnetic field which consumes a lot of time. Additional investigation is needed to adapt the method of micromagnetic simulations of IP demagnetized anisotropic domain states so it takes a reasonable amount of time. The magnetic structure of the Co(1) superlattice obtained by relaxation of the magnetic structure from the initial state of two skyrmions in zero magnetic field is shown in Fig. 9(c) for comparison. The average period of the structure is 500 nm which is larger than in the case of Fig. 9(b) and is consistent with the experimental results.

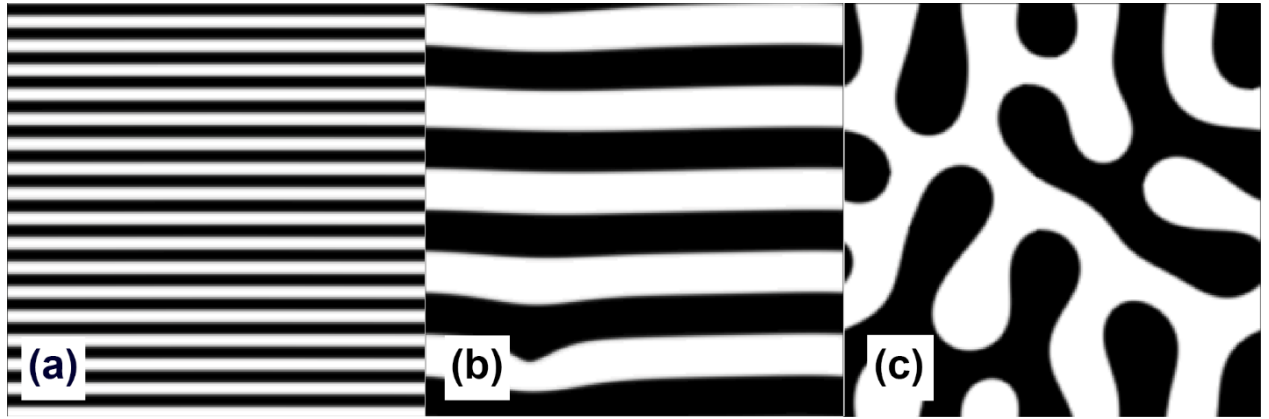

FIG. 9. Magnetic structures simulated in IP magnetic fields gradually decreasing from 1 T to zero (a) from uniform initial state without defects, (b) with two frozen skyrmions with a diameter of 50 nm. (c) Magnetic structure simulated from the uniform magnetic state with the magnetization oriented OP and two oppositely magnetized skyrmions with a diameter of 100 nm in zero magnetic fields.

*avdavydenko@gmail.com